\documentclass[preprint,floats,aps,showpacs,epsf]{revtex4-1}
\usepackage{amssymb}
\usepackage{epsfig,amsfonts}
\usepackage[fleqn]{amsmath}
\usepackage{amsthm,amssymb}
\usepackage{graphicx}
\usepackage{hhline}

\newcommand{\be}{\begin{equation}}
\newcommand{\ee}{\end{equation}}
\newcommand{\bea}{\begin{eqnarray}}
\newcommand{\eea}{\end{eqnarray}}

\newcommand{\p}{\partial}
\newcommand{\s}{\sigma}

\newcommand{\rd}{\mbox{d}}
\newcommand{\ri}{\mbox{i}}
\newcommand{\re}{\mbox{e}}

\begin{document}
\title{Polar Phase of 1D Bosons with Large Spin.}

\author{ G. V. Shlyapnikov$^{1,2}$ and A.  M. Tsvelik$^3$.}
\affiliation{
\mbox{$^1$ Laboratoire de Physique Th\'eorique et Mod\'eles Statistiques, Universit\'e. Paris Sud, CNRS,} \\
\mbox{91405~Orsay, France} \\
\mbox{$^2$ Van der Waals-Zeeman Institute, Universitry of Amsterdam, Science Park 904, }\\
\mbox{1098 XH Amsterdam, The Netherlands}\\
\mbox{$^3$Department of  Condensed Matter Physics and Materials Science, Brookhaven National Laboratory, }\\
\mbox{Upton, NY 11973-5000, USA}}
\date{\today}

\begin{abstract}

Spinor ultracold gases in one dimension represent an interesting example of strongly correlated quantum fluids. They have a rich phase diagram and exhibit a variety of quantum
phase transitions. We consider a one-dimensional spinor gas of bosons with a large spin $S$. A particular example is the gas of chromium atoms ($S=3$), where the dipolar
collisions efficiently change the magnetization and make the system sensitive to the linear Zeeman effect. We argue that in one dimension the most interesting effects come from the pairing 
interaction. If this interaction is negative, it gives rise to  a (quasi)condensate of singlet bosonic pairs with an algebraic order at zero temperature, and for $(2S+1)\gg 1$ the saddle point approximation leads to physically transparent results. Since in one dimension one needs a finite energy to destroy a pair, the spectrum 
of spin excitations has a gap. Hence, in the absence of magnetic field there is only one gapless mode corresponding to phase fluctuations of the pair quasicondensate. Once the 
magnetic field exceeds the gap another condensate emerges, namely the quasicondensate of unpaired bosons with spins aligned along the magnetic field. The spectrum then contains 
two gapless modes corresponding to the singlet-paired and spin-aligned unpaired bose-condensed particles, respectively. At T=0 the corresponding phase transition is of the 
commensurate-incommensurate type. 

\end{abstract}

\maketitle

\section{Introduction}

Spinor Bose gases attracted a great deal of attention in the last decade as they exhibit a much richer variety of macroscopic quantum phenomena than spinless bosons (see \cite{Ueda1} for
review). The physics of three-dimensional spin-1 and spin-2 bosons is rather well investigated, both theoretically \cite{Ho,Ohmi,Ueda2,Demler,Rizzi,Ueda3,Song,Turner} and in
experiments with Na and $^{87}$Rb atoms \cite{Ketterle,Sengstock,Chapman,Stamper-Kurn,Bloch}. The structure of the ground state strongly depends on the interactions, and in
particular ferromagnetic, polar (singlet-paired), and cyclic phases have been analyzed on the mean field level and beyond the mean field \cite{Ueda1}. The spinor physics of 3D
spin-3 bosons is described in Ref.~\cite{Santos} and, after successful experiments with Bose-Einstein condensates of $^{52}$Cr atoms ($S=3$) \cite{Pfau}, experimental studies of the spinor
physics in this system are expected in the near future.  

The observation of non-ferromagnetic states requires very low and stable magnetic fields (well below 1 mG) at which the interaction energy per particle exceeds the Zeeman energy.
Presently, the obtained stable field on the level of $0.1$ mG is expected to reveal a transition between ferromagnetic and non-ferromagnetic states in chromium
\cite{Bruno1}, and experiments using the magnetic field shielding and aiming at even lower fields are underway \cite{Gerbier}. 

It is important to emphasize that a change of magnetization of an atomic spinor gas under variations of the magnetic field requires spin-dipolar collisions, since the short-range
atom-atom interaction does not change the total spin. In dilute gases of sodium and rubidium the spin-dipolar collisions are very weak, and the magnetization does not feel a
change in the magnetic field on the time scale of the experiment. On the contrary, in a gas of chromium atoms which have a large magnetic moment of $6\mu_B$, the spin-dipolar
collisions efficiently change the magnetization and the gas becomes sensitive to the linear Zeeman effect \cite{Bruno2}.

Spinor Bose gases in one dimension (1D) are in many aspects quite different from their 2D and 3D counterpats and represent an interesting example of strongly correlated quantum
fluids. In this paper, having in mind the gas of chromium atoms ($S=3$), we assume that the system is sensitive to the linear Zeeman effect. We consider a 1D spinor gas of bosons 
where the dominant interactions are the density-density and the attractive pairing interactions. This choice is justified by the fact that in 1D only the latter interaction gives 
rise to a nontrivial quasi-long-range order. In contrast to 2D and 3D, in one dimension pairs with nonzero spin ${\bar S}$ do not condense. This is related to the fact that for 
${\bar S}\neq 0$ the symmetry of the condensate order parameter is non-Abelian. It is well known that strong quantum fluctuations in 1D dynamically 
generate spectral gaps for non-Abelian Goldstone modes which leads to exponential decay of the correlations (see, for example, \cite{polwieg}). As far as the polar phase (the 
condensate of ${\bar S}=0$ pairs) is concerned, it can be formed because the symmetry of the order parameter is Abelian. However, in 1D its  magnetic spectrum is quite different 
from that in 2D and 3D: in the absence of magnetic field the spin excitations have a gap. For a large spin $S$, the saddle point approximation gives a 
physically transparent description of the polar phase. A sufficiently large magnetic field closes the gap and leads to the transition from the singlet-paired (polar) phase to the 
ferromagnetic state. The presence of the spin-gap strongly changes the physics of the 1D polar phase and the polar-ferromagnetic transition compared to higher dimensions discussed 
for spin-3 bosons in Ref.~\cite{Santos}. We investigate the 1D polar phase and this quantum transition and discuss prospects for their observation in chromium experiments.
    
\section{The model}

As the atom-atom short-range interaction conserves the total spin, the Hamiltonian of binary interactions for (1D) bosons with spin $S$ can be written as a sum of projection operators 
on the states with different even spins ${\bar S}$ of interacting pairs \cite{Ueda1}:
 \bea
 V = \frac{1}{2}\int \rd x \sum_{{\bar S}=0}^{2S}\gamma_{\bar S}\hat{\cal P}_{\bar S}(x),  \label{V}
 \eea
where $x$ is the coordinate. For the 1D regime obtained by tightly confining the motion of particles in two directions, the interaction constants $\gamma_{\bar S}$ are related to
the 3D scattering lengths $a_{3D}(\bar S)$ at a given spin ${\bar S}$ of the colliding pair. Omitting the confinement induced resonance \cite{Olshanii} we have:
\bea
\gamma_{\bar S}=\frac{2\hbar^2}{Ml_0^2}\,a_{3D}(\bar S),    \label{gammaa}
\eea 
where $l_0=(\hbar/M\omega_0)^{1/2}$ is the confinement length, $M$ is the atom mass, and $\omega_0$ the confinement frequency.

 Imagine that all $\gamma_{\bar S}$ are equal to each other ($\gamma_{\bar
S}=\gamma>0$), except for $\gamma_{\bar S}$ at ${\bar S}=q$. We then use the relation $\sum_{\bar S}\hat{\cal P}_{\bar S}(x)=:\hat n^2(x):$ where $\hat n$ is the density operator
and the symbol $::$ denotes the normal ordering, and reduce the interaction Hamiltonian to the form $V=(1/2)\int dx (\gamma :n^2(x):+(\gamma_q-\gamma)\hat{\cal P}_q(x)$. For a
positive value of $(\gamma_q-\gamma)$ the system is an ordinary Luttinger liquid, but for $(\gamma_q-\gamma)<0$ the situation may change. In 3D a negative value of
$(\gamma_q-\gamma)$ would lead to a spontaneous symmetry breaking with a formation of the order parameter in the form of a condensate of pairs with  total spin $q$. In one
dimension only a quasi-long-range order is possible and only if $q=0$ when the symmetry  in question is an Abelian one \cite{polwieg}. Therefore, interactions with negative coupling constants, which
have $q$ different from zero or from $2S$ will not produce quasi-long-range-order. The case of $q =2S$ is exceptional because it corresponds to a ferromagnetic
state where the order parameter (the total spin) commutes with the Hamiltonian. Therefore, at $T=0$ this state can exist even in 1D. We do not discuss this interesting state, and the only possibility that remains is $q=0$.  So, in our model we have a (repulsive) density-density interaction and the pairing interaction that gives rise  to the formation of singlet pairs. 

In realistic systems the coupling constants $\gamma_{\bar S}$ are not equal to each other, although they are generally of the same order of magnitude.  We thus have to single out
the density-density interaction in a proper way and then deal with the rest. For example,  the interaction Hamiltonian (\ref{V}) can be represented  as a sum of squares of certain
local operators as is usually done in the theory of spinor Bose gases \cite{Ueda1,Santos}:
\bea
V= \frac{1}{2}\int \rd x\Big[c_0:{\hat n}^2(x) + c_1{\hat {\bf F}}^2(x) + c_2{\hat{\cal P}}_0(x) + c_3\mbox{Tr}{\hat{\cal O}}^2(x):+...\Big], \label{VCr}
\eea
where ${\bf F} = \psi^+{\bf S}\psi$, ${\cal O}_{ij} = \psi^+(S^iS^j)\psi$, the constants $c_i$  are linear combinations of $\gamma_{\bar S}$, and the symbol $...$ stands for
higher-order spin terms which we do not write. The operators $:\hat {\bf F}^2(x):$ and $:\hat{\cal O}^2(x):$ are given by $\sum_{\bar S}[{\bar S}({\bar S}+1)/2-S(S+1)]\hat{\cal
P}_{\bar S}(x)$ and  $\sum_{\bar S}[{\bar S}({\bar S}+1)/2-S(S+1)]^2\hat{\cal P}_{\bar S}(x)$ respectively, where the summation includes all values of ${\bar S}$ from zero to $2S$. We then
move the ${\bar S}=0$ part of these terms to the term $c_2\hat{\cal P}_0(x)$ and do the same procedure with higher order spin terms, which changes the constant $c_2$. The  $:\hat {\bf F}^2(x):$, $:\hat{\cal O}^2(x):$ etc. terms then no longer contain the interactions with ${\bar S}=0$ and, hence, can only lead to
renormalizations of the density-density and pairing interactions.

In the case of $^{52}$Cr we have  $c_0=0.65\gamma_6$, and the 3D scattering length is $a_6=112 a_B$ \cite{Santos}, where $a_B$ is the Bohr radius. The exact value of the 3D
scattering length $a_{3D}(0)$  is not known and, hence, the constants $\gamma_0$ and $c_2$ are also unknown. In this paper, when discussing $^{52}$Cr atoms we omit the 
$:\hat {\bf F}^2(x):$ and $:\hat{\cal O}^2(x):$ (renormalized) terms, treat $c_2$ as a free parameter and focus on the case of $c_2 < 0$. 

We then write down  the following Hamiltonian density in terms of the bosonic field operators $\Psi_j$:
 \bea
 {\cal H} =  \frac{1}{2M}\sum_j\p_x\Psi_j^+\p_x\Psi_j  + \frac{g}{2N}\Big[\sum_j\Psi^+_j\Psi_j\Big]^2 -
  \frac{g_0}{2N}\Big[\sum_j(-1)^j\Psi^+_j\Psi^+_{-j}\Big]\Big[\sum_j(-1)^j\Psi_j\Psi_{-j}\Big], \label{bare}
 \eea 
where the spin projection $j$ ranges from $-S$ to $S$, the coupling constant $g_0$ is assumed to be positive, and we put $\hbar=1$. The coupling constants $g_0$ and $g$ are related
to $c_0$ and $c_2$. For example, in the case of $^{52}$Cr we have $g = 7c_0=4.55\gamma_6>0$ and $g_0 = - c_2$.   

\section{Zero magnetic field. Saddle point approximation}

We now consider the case of $N =2S+1 >> 1$ and apply the $1/N$-approximation to the model described by the Hamiltonian density (\ref{bare}). First, we decouple the pairing from the density-density interaction by the Hubbard-Stratonovich transformation \cite{Hubbard}:
 \bea
 && - \frac{g_0}{2N}\Big[\sum_j(-1)^j\Psi^+_j\Psi^+_{-j}\Big]\Big[\sum_j(-1)^j\Psi_j\Psi_{-j}\Big] \rightarrow N|\Delta|^2/2g_0 + \Big[\Delta \sum_j (-1)^m\Psi_j^+\Psi_{-j}^+  +
h.c.\Big] \nonumber\\
 && \frac{g}{2N}\Big[\sum_j\Psi^+_j\Psi_j\Big]^2  \rightarrow N\lambda^2/2g + \ri\lambda\sum_j\Psi^+_j\Psi_j,
 \eea
where $\Delta(\tau,x)$ and $\lambda(\tau,x)$ are auxiliary dynamical fields. At large $N$ the path integral is dominated by the field configurations in the vicinity of the saddle
point $\Delta(\tau,x) = \Delta, ~~ \lambda(\tau,x) = \ri\lambda_0$. The values of $\Delta$ and $\lambda_0$ are determined self-consistently from the minimization of the free
energy. The stability of
the saddle point is guaranteed by the fact that the integration over the $\Psi,\Psi^+$ fields yields a term proportional to $N$ and therefore the entire action is $\sim N$. The
presence of large $N$ in the exponent in the path integral suppresses fluctuations of the fields $\Delta$ and $\lambda$, thus making the saddle point stable. 

 The bosonic action at the  saddle point is 
 \bea
 \tilde S = \sum_{\omega,k,m}(\Psi^+_{\omega,m}, \Psi_{-\omega,-m})\left(
 \begin{array}{cc}
 \ri\omega - \epsilon &(-1)^m \Delta\\
 (-1)^m\Delta^+ & -\ri\omega - \epsilon
 \end{array}
 \right)\left(
 \begin{array}{c}
 \Psi_{\omega,m}\\
 \Psi^+_{-\omega,-m}
 \end{array}
 \right), \label{MFS}
 \eea
 where $\epsilon = k^2/2M - \mu$ and $\mu = \mu_0 - \lambda_0$, with $\mu_0$ being the bare chemical potential.  From Eq.~(\ref{MFS}) we find the mean field spectrum of
quasiparticles (we assume that $\mu < 0$): 
 \bea
 E(k) = \frac{1}{2M}\sqrt{(k^2 + \kappa^2)(k^2 + k_0^2)}; ~~ \kappa^2 = 2M(-\mu - \Delta), ~~ k_0^2 = 2M(-\mu + \Delta). \label{QP}
 \eea
 The saddle point equations are: 
  \bea
  && \frac{\Delta}{g_0} = \int \frac{\rd\omega\rd k}{(2\pi)^2}\frac{\Delta}{\omega^2 + E^2(k)},\label{SP}\\
  && n = \int \frac{\rd\omega\rd k}{(2\pi)^2}\frac{\epsilon(k)-E(k)}{\omega^2 + E^2(k)},\label{SP1}\\
  && \mu = \mu_0 - gn, \label{mu}
  \eea
  where $n$ is the density of one of the bosonic species. 
  
 The quasiparticles (spin modes) constitute a $(2S+1)$-fold degenerate multiplet. As follows from Eq.~(\ref{QP}), the quasiparticles have a nonzero spectral gap 
   \bea
   E(0)\equiv m = k_0\kappa/2M . \label{gap}
   \eea
This result agrees with the one for $N=3$ obtained in Ref.~\cite{essler}. This is a special feature of one dimension. In 2D and 3D the integral in the saddle point equation
(\ref{SP}) does not diverge at small $\omega,k$ for $\kappa\rightarrow 0$, and such a gap is not formed. Therefore, one has a gapless spectrum of spin modes, which for $S=3,\,2$ and $1$ has been obtained 
in the studies of spinor Bose gases (see, e.g. \cite{Ueda1} and references therein). We would like to emphasize the fact that although Eqs.~(\ref{SP}),
(\ref{SP1}), and (\ref{mu}) resemble the equations for a superconductor, due to the bosonic nature of the problem the order parameter amplitude $\Delta$  is not equal to the
spectral gap, and the latter is related to the parameter $\kappa$.  
   
\begin{figure}
\centering
\includegraphics[width=8.5cm,angle=-90]{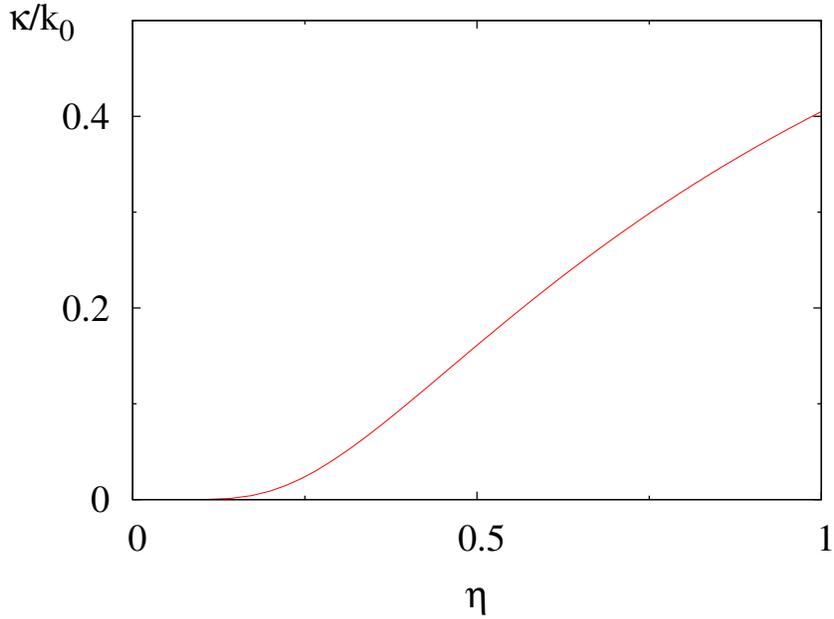}
\caption{The ratio $\kappa/k_0$  as a function of $\eta$. }
\label{fig1}
\end{figure}
  
\begin{figure}
\centering
\includegraphics[width=8.5cm,angle=-90]{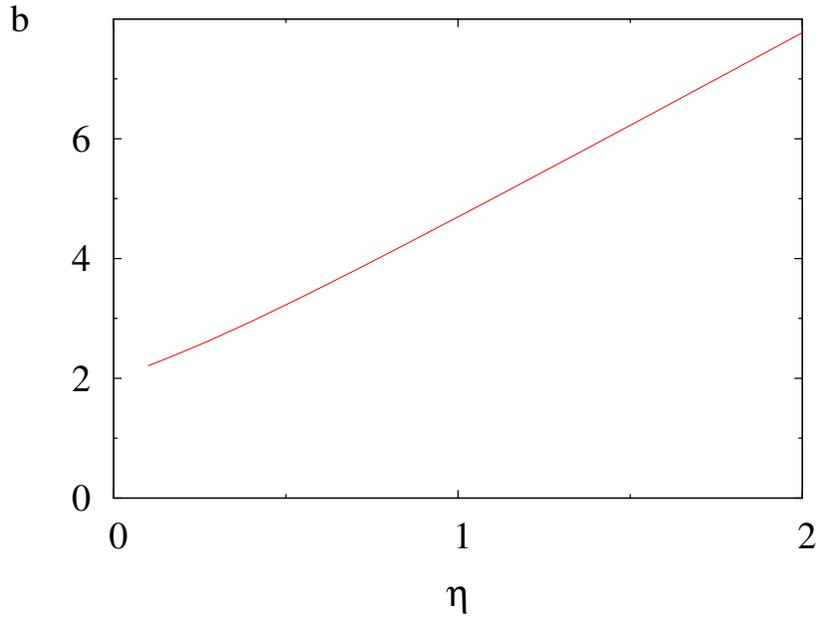}
\caption{The parameter $b$  as a function of $\eta$. }
\label{fig2}
\end{figure}

\begin{figure}
\centering
\includegraphics[width=8.5cm,angle=-90]{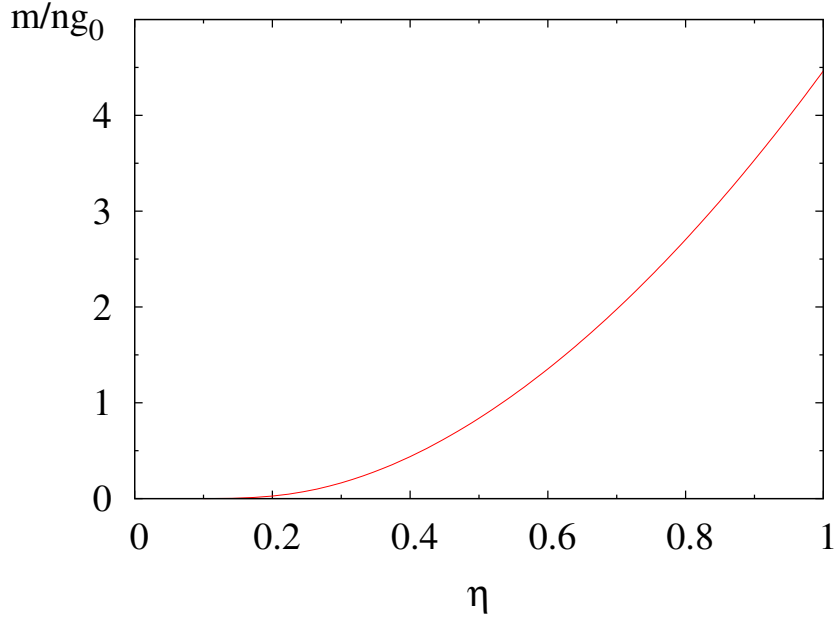}
\caption{The gap $m$ in units of $ng_0$ as a function of $\eta$. }
\label{fig3}
\end{figure}

  After the integration in Eqs.~(\ref{SP}) and (\ref{SP1}) we get the saddle point equations in the parametric form:
  \bea
  && 2\pi n = \frac{\kappa^2 + k_0^2}{2k_0}K\Big(\sqrt{1- \frac{\kappa^2}{k_0^2}}\Big) - k_0E\Big(\sqrt{1- \frac{\kappa^2}{k_0^2}}\Big)\\
  && \frac{\pi k_0}{Mg_0} = K\Big(\sqrt{1- \frac{\kappa^2}{k_0^2}}\Big), 
  \eea
  where $K(x),E(x)$ are elliptic functions \cite{ellipt}. From the form of these equations it is clear that the ratio $\kappa/k_0$ is a function of the parameter
\bea         \label{eta}
\eta=\frac{1}{2\pi}\left(\frac{Mg_0}{n}\right)^{1/2},
\eea
and $k_0$ can be written in the form: 
  \bea
  k_0 =b(\eta)(nMg_0)^{1/2}. \label{k0}
  \eea
Accordingly, Eq.~(\ref{gap}) for the gap takes the form:
\begin{equation}     \label{gapeta}
m=\frac{ng_0}{2}b^2(\eta)\frac{\kappa}{k_0}(\eta),
\end{equation}
so that the gap in units of $ng_0$ depends only on the parameter $\eta$. 

In the limit of weak interactions where $\eta\ll 1$, we obtain:
  \bea           \label{kappabweak}
\frac{\kappa}{k_0}\simeq\frac{4}{e}\exp\left(-\frac{1}{\eta}\right);\,\,\,\,\,b\simeq 2\label{kappa}
   \eea
and Eq.~(\ref{gap}) gives an exponentially small gap:
\bea
m\simeq \frac{8ng_0}{e}\exp\left(-\frac{1}{\eta}\right).  \label{gapweak} 
\eea   

For strong interactions, $\eta\gg 1$, we have
\bea          \label{b}     \label{kappabstrong}
\frac{\kappa}{k_0}\simeq 1;\,\,\,\,\,b\simeq \pi\eta,
\eea
and the gap is given by
\bea       \label{gapstrong}
m\simeq \frac{\pi^2\eta^2}{2}\,ng_0.
\eea

The numerically obtained dependence of $\kappa/k_0$ is displayed in Fig.~1, and the function $b(\eta)$ is shown in Fig.~2. The gap is presented in Fig.~3.
The asymptotic formula (\ref{gapweak}) obtained in the limit of small $\eta$ already works with 20$\%$ of accuracy for $\eta=0.05$. With the same accuracy the large-$\eta$ 
asymptotic formula (\ref{gapstrong}) is already valid for $\eta=1$.

In the limit of weak interactions, taking into account that $|\mu| \approx \Delta$ and using Eqs.~(\ref{QP}) and (\ref{k0}), we get $\mu \approx -ng_0/4M$. Substituting this
relation into Eq.~(\ref{mu}) we obtain
  \bea
  n = \frac{\mu_0}{g-g_0}. \label{chem}
  \eea
  Hence the system is thermodynamically stable for $g > g_0$.

   The only gapless excitation of the system is the phase mode of the complex scalar field $\Delta$. This excitation describes sound waves of the pair condensate. The effective
Hamiltonian  for the phase mode $\Phi$ is
  \bea
  H_{phase} = \frac{v}{2}\int \rd x \Big[K_s\Pi^2 + K_{s}^{-1}(\p_x\Phi)^2\Big], ~~ [\Pi(x),\Phi(y)] = -\ri\delta(x-y),
  \eea
  where $\Pi$ is a canonically conjugate momentum. The velocity $v$ and Luttinger parameter $K_s$ are extracted from the functional derivatives of the saddle point action and are given by the following equations:
  \bea
  && (K_sv)^{-1} = -N\Delta^2\int \frac{\rd\omega\rd k}{2(2\pi)^2}\frac{\p G(\omega,k)}{\p\omega}\frac{\p G(-\omega,-k)}{\p\omega} = \nonumber\\
  && =N\Delta^2\int \frac{\rd\omega\rd k}{2(2\pi)^2}\frac{(\epsilon^2 + \omega^2 - \Delta^2)^2 +(2\Delta\omega)^2}{(\omega^2 + \epsilon^2 - \Delta^2)^4}= \frac{N\Delta^2}{8\pi}\int
\rd k \frac{2E^2 + \Delta^2}{E^5}= \nonumber\\
  && \frac{N\Delta^2}{4\pi}\frac{M^3}{\kappa^5}f_1(k_0/\kappa), \\
  && v/K_s = -N\Delta^2\int \frac{\rd\omega\rd k}{2(2\pi)^2}\frac{\p G(\omega,k)}{\p k}\frac{\p G(-\omega,-k)}{\p k} = \nonumber\\
  && N\Delta^2\int \frac{\rd\omega\rd k}{(2\pi)^2}(k/M)^2\frac{(\epsilon^2 + \omega^2 - \Delta^2)^2 +(2\Delta\epsilon)^2}{(\omega^2 + \epsilon^2 - \Delta^2)^4}=  \frac{N\Delta^2}
{8\pi}\int \rd k (k/M)^2\frac{2E^4 + 5(\Delta\epsilon)^2}{E^7} = \nonumber\\
  && \frac{2N\Delta^2}{\pi}\frac{M}{\kappa^3}f_2(k_0/\kappa)
  \eea
  where $G(\omega,k)$ is the Green function of the field $\Psi_j$, defined as $\langle\langle\Psi_j(\omega,k)\Psi^+_{j'}(\omega,k)\rangle\rangle=\delta_{jj'}G(\omega,k)$, and 
  \[
  f_1(x) = \int_0^{\infty} \rd y \frac{8(1+y^2)(y^2 +x^2) + (1-x^2)^2}{[(y^2+1)(y^2 +x^2)]^{5/2}}, 
  \]
  \[
  f_2(x) = \int_0^{\infty}\rd y y^2\frac{2(y^2+1)^2(y^2+x^2)^2 + (5/4)(y^2 + x^2/2 + 1/2)^2(x^2-1)^2}{[(y^2+1)(y^2 +x^2)]^{7/2}}.
  \]
The functions $f_1(k_0/\kappa)$ and $f_2(k_0/\kappa)$ can be expressed in terms of elliptic functions $E(\sqrt{1-\kappa^2/k_0^2})$ and $K(\sqrt{1-\kappa^2/k_0^2})$, but the
expressions are combersome and we do not present them. In the limit of weak interactions we have:
\bea      \label{f12weak}
f_1(k_0/\kappa)=\frac{2\kappa}{3k_0};\,\,\,\,\,f_2(k_0/\kappa)=\frac{k_0}{24\kappa};\,\,\,\,\,\eta\ll 1\,\,\, {\rm and}\,\,\,\kappa\ll k_0,
\eea
whereas for strong interactions these functions are given by
\bea      \label{f12strong}
 f_1(k_0/\kappa)=\frac{3\pi}{2};\,\,\,\,\,f_2(k_0/\kappa)=\frac{\pi}{8};\,\,\,\,\,\eta\gg 1\,\,\, {\rm and}\,\,\, \kappa\simeq k_0.
\eea
  The velocity $v$ is given by the relation: 
  \bea
  v = \sqrt 8\frac{k_0}{M}\left\{\frac{\kappa}{k_0}\left[\frac{f_2(k_0/\kappa)}{f_1(k_0/\kappa)}\right]^{1/2}\right\}, 
  \eea
so that for weak interactions using Eqs.~(\ref{f12weak}), (\ref{k0}) and $b\simeq 2$ we have:
\bea         
v\simeq \sqrt{\frac{2ng_0}{M}};\,\,\,\,\,\,\eta\ll 1.
\eea
For strong interactions equations (\ref{f12strong}), (\ref{k0}), and (\ref{kappabstrong}) lead to
\bea
v\simeq \sqrt{\frac{2ng_0}{3M}}\,\pi\eta;\,\,\,\,\,\,\,\eta\gg 1.
\eea
The dependence of $v$ on the parameter $\eta$ is displayed in Fig.~4.

The Luttinger parameter $K_s$ follows from the relation:
  \bea
  K_s^{-1} = \frac{N}{16\pi}\left(\frac{k_0^2}{\kappa^2}-1\right)^2\sqrt{f_1(k_0/\kappa)f_2(k_0/\kappa)}.
  \eea
 The scaling dimension of the $\Delta$ field is $d= K_s/4\pi$  and it decreases very rapidly with $\eta$, which indicates that the mean field approximation works very well.
In Fig.~5 we show the dependence of $d$ on $\eta$ for $N=7$. In the regime of weak interactions the scaling dimension is exponentially small and it remains significantly smaller
than unity even for $\eta\simeq 2$.

  
  \begin{figure}
\centering
\includegraphics[width=10cm,angle=-90]{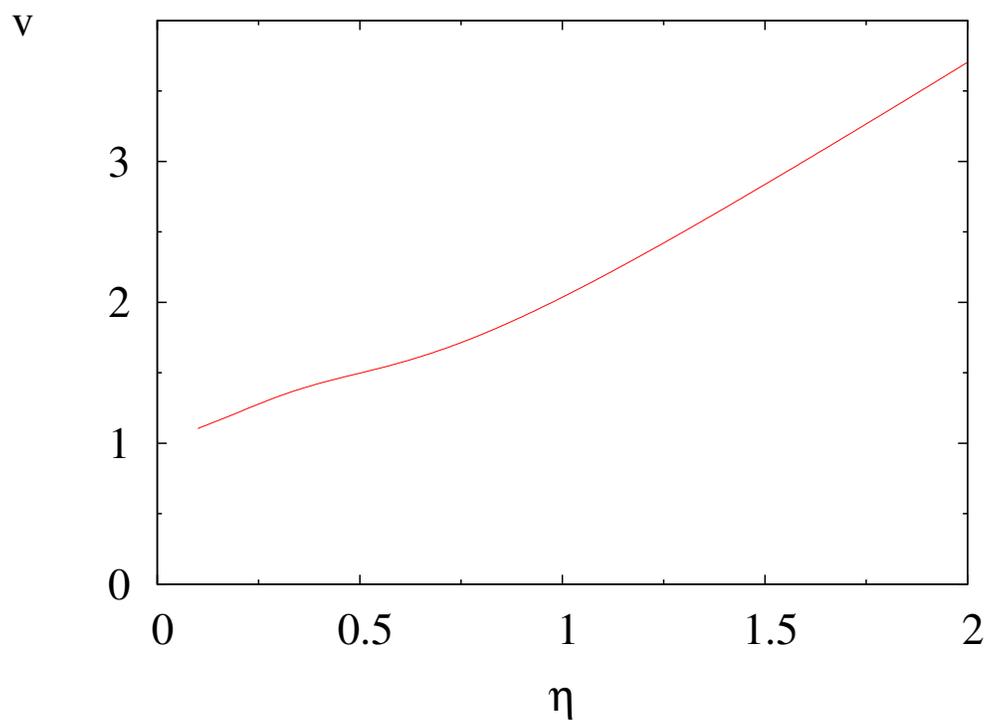}
\caption{The velocity $v$ in the units of $\sqrt{2ng_0/M}$ as a function of $\eta$.}
\label{fig-v}
\end{figure}

  \begin{figure}
\centering
\includegraphics[width=10cm,angle=-90]{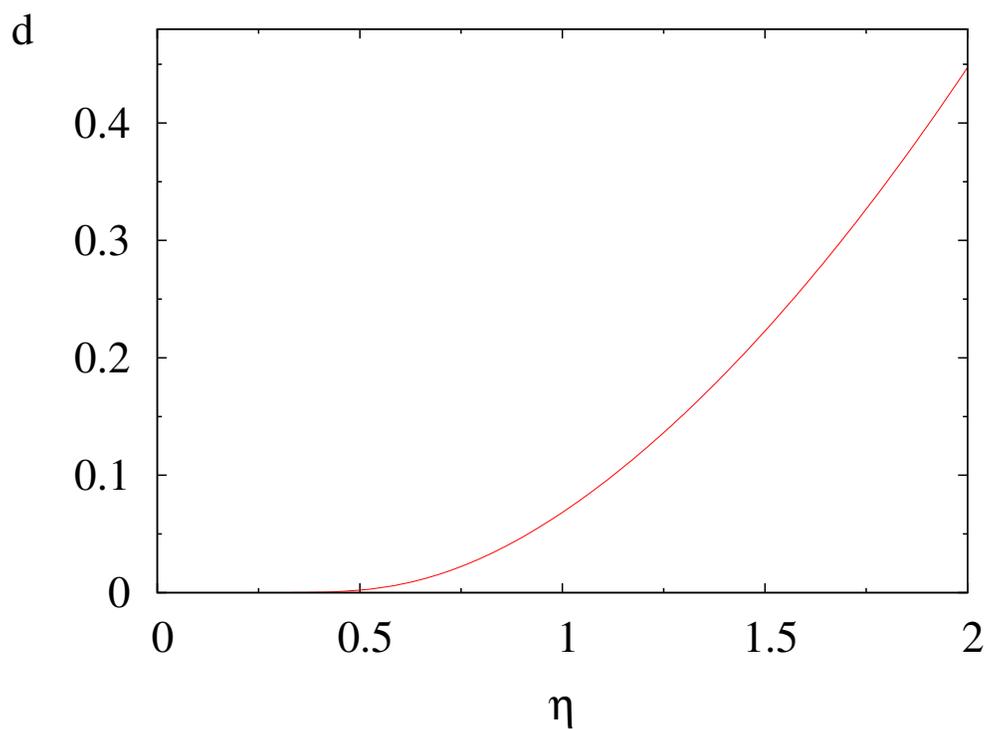}
\caption{The scaling dimension $d$ of the order parameter field $\Delta$  for $N=7$ as a function of $\eta$.}
\label{fig-D}
\end{figure}

To conclude this part we give a brief summary of the properties of the paired phase. With certain modifications, the properties for an arbitrary large spin $S$ are similar to the
ones for $S=1$ described in Ref.~\cite{essler}. Namely, all single particle correlation functions decay exponentially. This follows from the fact that the operator $\psi^+_j$
always emits a gapped vector excitation (Bogolyubov quasiparticle) from the $(2S+1)$-fold degenerate multiplet (for $S=1$ it is a gapped triplet). Two-particle correlation
functions of the operators $\psi_j\psi_j$ and their Hermitean conjugates (no summation assumed) undergo a power law decay.

 \section{Magnetic Field and Exact Solution}
 
In order to study the influence of the magnetic field on the properties of the singlet-paired phase one can also use the saddle point approximation employed in the previous
section. However, the saddle point equations become too involved. Therefore we resort to non-perturbative methods. 
  
 As was demonstrated in Ref.~\cite{jiang}, the model described by the Hamiltonian density (\ref{bare}) possesses U(1)$\times$O(2S+1) symmetry. Therefore it is reasonable to suggest
that the low energy sector of this model is
described by a combination of the U(1) Gaussian theory and the O(2S+1) nonlinear sigma (NL$\s$) model. For  S=1 this was explicitly demonstrated in Ref.~\cite{essler}. Both the
U(1) theory and the sigma model are integrable and the exact solution gives access to the low energy sector of the model. At a special ratio of the coupling constants one can get
even further, since it was demonstrated \cite{jiang} that the entire model (\ref{bare}) is integrable at a particular ratio of $g_0/g$.  Below we restrict our consideration to the
low energy sector where we are not constrained to this particular ratio. 
  
 As we have said, the O(N) NL$\s$ model is exactly solvable. In the absence of magnetic field its excitations are massive particles transforming under the vector representation of
the O(N) group. This agrees with our result for model (\ref{bare}) based on the large $N$ approximation. As is always the case for Lorentz invariant integrable models, all the
information on the thermodynamics is contained in the two-body S-matrix which was found in Ref.~\cite{zamzam}. Consider $N=2S+1$ ($S$ integer) and physical particles which have a
relativistic-like spectrum $\epsilon(\theta) =m\cosh\theta, ~~p(\theta) = \tilde v^{-1}m\sinh\theta$, with mass (gap) $m$, velocity $\tilde v$, and the total energy  
  \bea
  E = m\sum_i \cosh\theta_i.
  \eea
For the particles  confined in a box of length $L$ with periodic boundary conditions, the Bethe Ansatz equations read:  
  \bea
  && \exp[\ri \tilde v^{-1}mL\sinh\theta_i] = \prod_{i\neq j}^nS_0^{-1}(\theta_i -\theta_j)\prod_{a_1}^{m_1}\frac{\theta_i - \lambda_{a_1}^{(1)} - \ri\pi}{\theta_i - \lambda_{a_1}^{(1)} +
\ri\pi},\nonumber\\
  &&\prod_{i=1}^n\frac{\lambda_{a_1}^{(1)} -\theta_i - \ri\pi}{\lambda_{a_1}^{(1)}- \theta_i + \ri\pi}\prod_{a_2=1}^{m_2}\frac{\lambda_{a_1}^{(1)} -\lambda_{a_2}^{(2)} - \ri\pi}
{\lambda_{a_1}^{(1)}- \lambda_{a_2}^{(2)}+ \ri\pi} = \prod_{b_1=1}^{m_1}\frac{\lambda_{a_1}^{(1)} -\lambda_{b_1}^{(1)} - 2\ri\pi}{\lambda_{a_1}^{(1)}- \lambda_{b_1}^{(1)}+
2\ri\pi},\nonumber\\
  && ...\nonumber\\
  && \prod_{a_{S-1}=1}^{m_{S-1}}\frac{\lambda_{a_S}^{(S)} -\lambda_{a_{S-1}}^{(S-1)} - \ri\pi}{\lambda_{a_1}^{(1)}- \lambda_{a_2}^{(2)}+ \ri\pi} =
\prod_{b_S=1}^{m_S}\frac{\lambda_{a_S}^{(S)} -\lambda_{b_S}^{(S)} - \ri\pi}{\lambda_{a_S}^{(S)}- \lambda_{b_S}^{(S)}+ \ri\pi}, \label{Bethe}
  \eea
  where $S_0$ is related in the standard way to the integral kernel $K(\theta)$:
  \[
  \frac{1}{2\ri\pi} \frac{\rd \ln S_0(\theta)}{\rd\theta}  = \delta(\theta) - K(\theta),
  \]
  and
 \[
  K(\theta) = \int \re^{\ri\omega\theta/\pi} K(\omega)\rd \omega/2\pi, ~~ K(\omega) = \frac{1- \exp[-2|\omega|/(N-2)]}{1+\exp[-|\omega|]}.
  \]
 
The spectral gap $m$ (the particle mass) and velocity $\tilde v$ are related to the bare parameters of the model (\ref{bare}). For $N\gg 1$ one can use Eq.~(\ref{gapeta}) for the gap, and in the 
low-energy limit the spectrum (\ref{QP}) of the spin modes becomes $E(k)=\sqrt{m^2+\tilde v^2k^2}$ with $\tilde v=(b/2)\sqrt{ng_0(1+\kappa^2/k_0^2)/M}$, so that in the limit of weak interactions 
we have $\tilde v \approx \sqrt{ng_0/M}$.

  Equations (\ref{Bethe}) constitute a system of $S+1$ coupled algebraic equations for the quantities $\theta_1,...\theta_n, ~~\lambda_1^{(1)},...\lambda_{m_1}^{(1)} ~~
\lambda_1^{(2)},...\lambda_{m_2}^{(2)}, ... ~~ \lambda_1^{(S)},...\lambda_{m_S}^{(S)}  $.  Integer numbers $m_i$ are  eigenvalues of the $S$ Cartan generators of the group O(2S+1). 
From Eq.~(\ref{Bethe}) it is obvious that the total energy of the system depends on configurations of $\lambda$'s and through them it depends on spin indices of constituent
particles. Some of the Cartan generators for the problem under consideration were constructed in Ref.~(\cite{jiang}) where it was also shown that the projection of the total spin
of the system is
  \be
  S^z = Sn - \sum_i m_i. \label{Sz}
  \ee
The Cartan generators commute with the Hamiltonian, and $m_i$'s are integrals of motion. Therefore, the magnetic field which couples to $m_i$ through $S^z$ (\ref{Sz}), does not
violate integrability. Moreover, once the field is applied, the energies of all eigenstates with $m_i \neq 0$
go up. Thus, at sufficiently low temperature one may consider only eigenstates with  no $\lambda$-rapidities, since their energies decrease in the field. When the magnetic field
exceeds the spectral gap $m$, the $\theta$-rapidities  start to condense creating a Fermi sea. In the ground state the $\theta$-rapidities are distributed over a finite interval
$(-B,B)$. The distribution function $\rho(\theta)$  and magnetization $S^z$ are determined by the following integral equations:
  \bea
  && \int_{-B}^B K(\theta - \theta')\rho(\theta')\rd \theta' = \frac{m}{2\pi \tilde v}\cosh\theta, \label{dens}\\
  && \int_{-B}^B K(\theta - \theta')\epsilon(\theta')\rd \theta' = m\cosh\theta - hS, ~~ \epsilon(\pm B) = 0\label{En}\\
  && S^z/L = S\int_B^B\rd\theta \rho(\theta), 
  \eea
  where $h = g_L\mu_B{\cal B}$, with ${\cal B}$ being the magnetic field and $g_L$ the Landee factor. There is obviously one transition  at $h_c = m/S$. The magnetization is zero
for $h<h_c$ and it gradually increases with the field for $h>h_c$ (there is another transition in high magnetic fields corresponding to the saturation of the
magnetization, but the low energy theory cannot describe it). In order to find the magnetic field dependence of the magnetization near the transition, where $B << 1$, we
approximate
the kernel as
  \bea
  K(\theta) = \delta(\theta) - A + O(\theta^2), ~~ A= \ln 4 + \psi(1/2 + 2/N-2) - \psi(2/N-2),
  \eea
  and look for the solution of Eqs.~(\ref{dens}) and (\ref{En}) in the form:
  \bea
  \epsilon(\theta) \approx a(\theta^2 -B^2), ~~ \rho(\theta) \approx \mbox{const}. \label{epsrho}
  \eea
  Substituting  $\epsilon(\theta)$ and $\rho(\theta)$ given by Eq.~(\ref{epsrho}) into Eqs.~(\ref{dens}) and (\ref{En})) we get:
  \bea
  \!\!\!\!\frac{S^z}{L} = \frac{S}{\pi\tilde v}\Big[2m(hS\!-m)\Big]^{1/2}\Big\{1\!+\!8A/3[2(hS-m)/m]^{1/2} + O((hS-m)/m)\Big\};\,h>h_c.   \label{Sz1}
  \eea
Keeping only the leading term in Eq.~(\ref{Sz1}) and restoring the dimensions we have:
\bea        \label{Sz2}
\frac{S^z}{L}=\frac{\sqrt{2}Sm}{\pi\hbar\tilde v}\left(\frac{{\cal B}-{\cal B}_c}{{\cal B}_c}\right)^{1/2};\,\,\,\,\,\,{\cal B}>{\cal B}_c,
\eea
with the critical magnetic field given by 
\bea         \label{Bc}
{\cal B}_c=\frac{m}{g_L\mu_B S}.
\eea
Equation (\ref{Sz2}) shows a typical field dependence of the magnetization for the quantum commensurate-incommensurate transition. This transition was first studied by Japaridze and Nersesyan \cite{japners} in the context of spin systems with a gap, where (as in our case) it is driven by the magnetic field. Later Pokrovsky and Talapov considered such a transition in the charge sector, where it is driven by a change in the chemical potential \cite{pokr}. The magnetization is exactly zero below the critical field and increases as $\sqrt{{\cal B}-{\cal B}_c}$ above the critical field near the transition. Note that this is quite different from the 3D case where the magnetization decreases continuously with the magnetic  field when the latter goes below the critical value (see, e.g. \cite{Santos}).
  
  \section{Conclusions}
  
We have found that the phase diagram and general properties of 1D bosons with a large spin $S$ resemble the properties of spin-1 bosons. For the attractive pairing interaction 
($g_0>0$) the bosons with opposite spin projections create pairs which bose-condense giving rise to quasi-long-range order. The saddle point approximation based on the condition of
large $N=2S+1$, gives a transparent picture of the emerging polar phase. The peculiarity of one dimension is that all spin excitations have a spectral gap. Hence, in the absence of magnetic field
there is only one gapless mode corresponding to phase fluctuations of the pair quasicondensate. Once the magnetic field exceeds the gap,  another  quasicondensate emerges. This is
the condensate of unpaired bosons with spins aligned along the magnetic field. The spectrum then acquires two gapless modes corresponding to the singlet-paired and spin-aligned
unpaired bosons, respectively. At T=0 the corresponding phase transition is of the commensurate-incommensurate type, which is qualitatively similar
to what we have in the case of the O(3) NL$\s$ model. There is a second transition at high magnetic fields corresponding to the saturation of
the magnetization. However, it is not described by the low-energy theory and is beyond the scope of this paper. In the context of ultracold quantum gases, the commensurate-incommensurate transition has also been discussed for one-dimensional spin-$3/2$ fermions in the presence of the quadratic Zeeman effect \cite{Santos2}.

The observation of the commensurate-incommensurate phase transition in a 1D gas of $^{52}$Cr atoms would require (aside from a positive value of $g_0$ and, hence, a negative $c_2$) fairly
strong interactions corresponding to the parameter $Mg_0/{\bar n}\sim 1$, where ${\bar n}=7n$ is the total density, so that $\eta\sim 0.5$. Then the spin gap in the polar phase is
of the order of $ng_0$ and (assuming $g_0$ by a factor of 3 smaller than $g$) can be made on the level of 100 nK at 1D densities ${\bar n}\sim 10^5$ cm$^{-1}$. Then 
the transition occurs at the critical field of the order of $0.2$ mG and can be observed at temperatures $T\sim 20$ nK. This however is likely to require the 1D regime with a
rather strong confinement in the transverse directions (with a frequency of the order of 100 kHz as in the ongoing chromium experiment in the 1D regime at Villetaneuse
\cite{Bruno1}).

{\it Note added}

After this work has been finished the Villetaneuse group has reported the observation of the demagnetization transition for $^{52}$Cr atoms in the 1D regime, under a decrease of the 
magnetic field to below $0.5$ mG. However, the experiment is done at temperatures $\sim 100$ nK and the state which is reached by decreasing the magnetic field does not necessirily 
reveal the nature of the ground state due to thermal excitations (and due to diabaticity at the transition in the experiment).

   \section{Acknowledgements}
We are grateful to B. Laburthe-Tolra, P. Pedri, and L. Santos for fruitful discussions and acknowledge valuable remarks of A. Georges. This work was supported by  the US Department
of Energy, Basic Energy Sciences, Material Sciences and Engineering Division, by the IFRAF Institute of Ile de France, by ANR (Grant 08-BLAN0165), and by the Dutch Foundation FOM.
We also express our gratitude to the Les Houches Summer School ''Many-Body Physics with Ultracold Atoms'' for hospitality. LPTMS is a mixed research unit No. 8626 of CNRS and
Universit\'e Paris Sud.

\end{document}